\documentclass[traditabstract]{aa}
\usepackage{times}
\usepackage{graphics,graphicx,natbib}
\bibpunct{(}{)}{;}{a}{}{,}

\def\ltsima{$\; \buildrel < \over \sim \;$}
\def\lsim{\lower.5ex\hbox{\ltsima}}
\def\gtsima{$\; \buildrel > \over \sim \;$}
\def\gsim{\lower.5ex\hbox{\gtsima}}

\def\xmm{{\em XMM--Newton}}

\def\grb {GRB\,120711A}

\begin{document}

\title{A search for lines  in the bright X-ray afterglow of \grb\ }

\author{A. Giuliani\inst{1}, S. Mereghetti\inst{1}}

\institute{INAF, Istituto di Astrofisica Spaziale e Fisica Cosmica Milano, via E.\ Bassini 15, I-20133 Milano, Italy}

\offprints{}

\date{Received 02/04/2013 / Accepted 16/07/2013}

\authorrunning{A. Giuliani and S. Mereghetti}
\titlerunning{A search for lines in the afterglow of GRB 120711A}

\abstract { \grb , discovered and rapidly localized by the INTEGRAL satellite, attracted particular interest due to its 
high  {$\gamma$-ray} fluence, very bright X-ray afterglow, and the detection of a prompt optical transient 
and of long-lasting emission at GeV energies. A follow-up observation carried out with the \xmm\  satellite has provided an X-ray spectrum in the 0.3-10 keV with unprecedented statistics for a GRB afterglow 20 hours after the burst.
The spectrum is well fit by a power-law with photon index { 1.87$\pm$0.01,
modified by absorption in our Galaxy and in the GRB host  {at z=1.4}.
A Galactic absorption consistent with that estimated from neutral hydrogen observations is obtained only
with host metallicity lower than 5\% of the Solar value.}
We report the results of a sensitive search for emission and absorption lines using the matched filter smoothing method (Rutledge \& Sako 2003).
No statistically significant lines were found.
The upper limits on the equivalent width of emission lines, derived through Monte Carlo simulations, are few tens of eV, a factor $\sim$ 10 lower than that of the possible lines reported in the literature for other bursts.
}

\maketitle

\section{Introduction}

A few detections of X-ray emission and absorption lines in  $\gamma$-ray bursts (GRB) were reported in the first years after the discovery of GRB afterglows \citep{piro00,yoshida01,watson02,reeves03,butler03,mereghetti03},  based on data obtained with the $BeppoSAX$, $ASCA$, \xmm, and $Chandra$ satellites.
Although the  statistical significance of these  detections was never higher than 5$\sigma$, and more typically around 3$\sigma$, these results attracted considerable interest in view of the potential diagnostic for the physics of GRBs and the properties of their environments offered by these putative features \citep[ and references therein]{vietri01}.

After the launch of $Swift$,  it has become possible to explore the afterglows at earlier times, but a comprehensive analysis of 40  afterglows observed with the $Swift/XRT$ instrument failed to detect statistically significant lines \citep{hur08}.
A critical assessment of the possible line features in the nine afterglows reported before 2004 was presented in \citet{sak05}.
These authors performed  a uniform, systematic   analysis of the spectra obtained with the different instruments
and estimated the   significance of the lines with
Monte Carlo simulations. They also properly took into account the fact that, in the  {absence} of a known redshift,
the line energies should be considered as  free parameters. This increases the number of independent
searches and the detection significances have to be reduced accordingly.
In their analysis, \citet{sak05}   obtained lower significances for these lines, compared to those claimed in the previous papers.
Other   effects, e.g.,  the different continuum components adopted in the spectral fit, the estimates of the   background, and  imperfections in the instrumental response matrices, can also lead to  discordant results on the line detections significance.
For example, according to \citet{but05}, the lines from low-Z elements in GRB 011211 are at the $\sim$3$\sigma$ level  only
if the interstellar absorption is fixed at the Galactic value, while their significance decreases if the column density is allowed to vary.

Even if the statistical evaluation is done properly and all the technical issues are carefully taken into account, a great limitation to line searches comes from the small counting statistics of the available spectra. The instrument with the largest collecting area used so far in these studies is the EPIC camera on \xmm , but its observations are generally done when the afterglow flux has significantly decreased, since a few hours are needed to re-point this satellite after a GRB discovery.

Here we report on an \xmm\ observation of the very bright \grb\ which, providing an afterglow spectrum of unprecedented statistics, offers the possibility to carry out a very sensitive search for spectral features.

\begin{figure}
\includegraphics[width=65mm,angle=90]{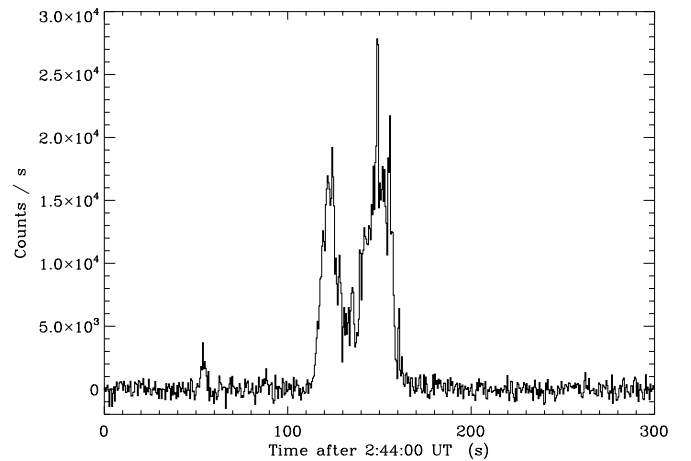}
\caption{\footnotesize {Light curve of \grb\ at E$>$80 keV in bins of 0.5 s measured with the anti-coincidence system of the SPI instrument. }
\label{fig_lc}}
\end{figure}

\section{ \grb\ }

\grb\  was  discovered  with the IBIS instrument \citep{ube03}  on the INTEGRAL satellite \citep{13434}.
Its light curve, obtained with the anti-coincidence system (ACS) of the INTEGRAL SPI instrument \citep{acs}, is shown in Fig.~\ref{fig_lc}.
The INTEGRAL Burst Alert System (IBAS, \citet{mer03}) triggered on the precursor peak that occurred at about 02:44:50 UT.
The rapid IBAS localization  allowed  the identification of an optical transient  \citep{13430} reaching a V band magnitude of $\sim$12  roughly 112 seconds after the trigger \citep{13433}, while the hard X-ray prompt emission was still active.

The burst had  a fluence of 5$\times$10$^5$  ACS counts, which corresponds to $\sim$5$\times$10$^{-5}$   erg cm$^{-2}$  in the 75 keV - 1 MeV energy range\footnote{The ACS does not provide spectral information; we adopted the average conversion factor derived
by \citet{vig09}.}.
This agrees with the values of a few 10$^{-4}$   erg cm$^{-2}$  (E$>$20 keV) measured by several other
instruments  \citep{13468,13446,13437}.

 \grb\ is among the bursts with the highest fluence ever detected \cite[see. e.g.][]{goldstein12,meegan96}.
Its time integrated spectrum is well described by a Band
function with typical parameters of hard GRBs, $\alpha$=--1,  $\beta$$\sim$--2.5, and  $E_{peak}\sim$1 MeV \citep{13437,13446}.

Also the X-ray afterglow of \grb\ was particularly bright. Immediately after the main burst, significant emission at energy above $\sim$15 keV
was detected for at least 1000 s in IBIS  \citep{13435} and for 800 s in the 20-100 keV range with SPI  \citep{13468}.
\textit{Swift/XRT} started to observe  the GRB position 2.3 hours  after the prompt emission and detected X-ray emission  with a light curve decaying as a power-law   with index  $1.59^{+0.16}_{-0.15}$  \citep{13442}.
The X-ray flux measured 11 hr after the burst was 2.96$\times10^{-11}$ erg cm$^{-2}$ s$^{-1}$.
Compared to all the other X-ray afterglows observed by \textit{Swift}, this is the third brightest ever, lying at about 3$\sigma$ from the average in the distribution of the afterglow fluxes measured with \textsl{Swift/XRT}.

 {A suggested  photometric redshift of z$\sim$3 \citep{13438} was not confirmed by optical spectroscopic
observations, which showed, instead, absorption lines  of MgII and FeII at z=1.405 \citep{13441}. 
In the following, we will assume this value for the redshift of \grb .}

\begin{figure*}[ht!]
\includegraphics[width=170mm]{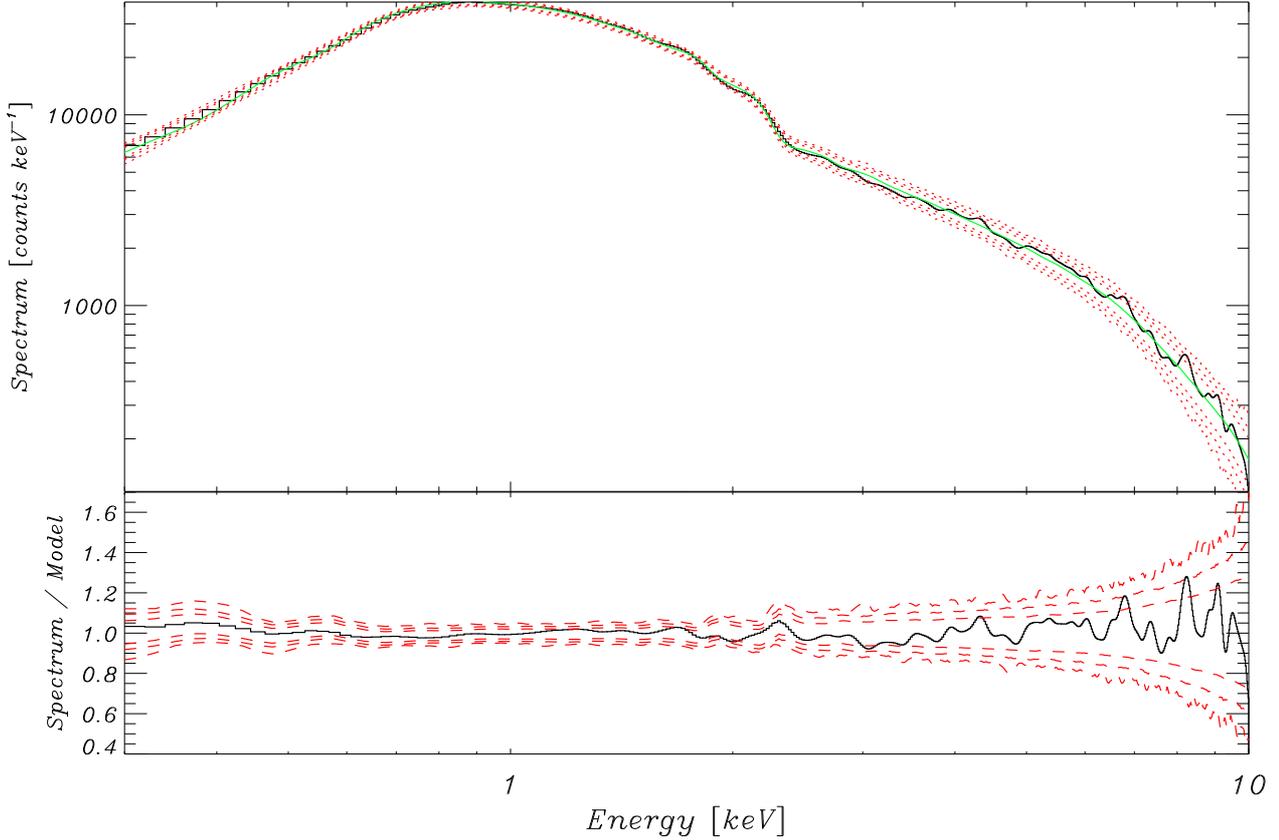}
\caption{\footnotesize {Upper panel: Spectrum of the X-ray afterglow of \grb\ convolved with a Gaussian as described in the text (black solid line).
The best-fit model is  shown by the green line. The dashed lines are the percentile curves  obtained with the simulations for emission and
absorption lines. They correspond to  single-trial detection significance of  2, 3 and 4$\sigma$.
Lower panel: the same curves of the upper panel, normalised by dividing by the best-fit model.}
 \label{fig_spectrum}}
\end{figure*}

\section{Observations and data reduction}

The \xmm\ satellite started to observe the afterglow of \grb\ on July 12, 2012  at 00:30 UT.
The observation lasted 52 ks.
Here we concentrate on the data obtained with the  EPIC pn instrument,  which consists of
a CCD camera operating in the 0.2-12 keV energy range and  with an energy resolution (FWHM) of $\sim$ 90 eV at 1.5 keV and $\sim$ 150 eV at 6 keV  \citep{str01}.
The instrument was operating in the Full Frame mode, giving a
time resolution of 73 ms,
and with the thin optical blocking filter.

We processed the EPIC data with version 12 of the \xmm\ Standard Analysis Software (SAS v12).
The data were cleaned by discarding a few time intervals of enhanced background, which resulted in a live time exposure  of 35.2 ks.
For  the source spectrum we used mono- and bi-pixel events (PATTERN $\leq$ 4) extracted from a circular  region with radius of 40$''$ and rebinned   to have at least 30 counts for each energy channel.
The background spectrum was  extracted from a source-free region  on the same chip as the target.
Spectral fits were done with the XSPEC 12.7.0 package. 
For the  {Galactic} interstellar absorption we used the \textit{phabs} model, with abundances from \cite{anders89}.

\section{Spectral analysis}

 {We fit the EPIC/pn spectrum in the 0.3-10 keV energy range with a power law model modified by interstellar absorption in our Galaxy, 
N$_{H}^{Gal}$, and in the GRB host,  N$_{H}^{Host}(z)$,  with  redshift fixed at $z$=1.405.
The best fit obtained using Solar abundances in the host absorption resulted in a low    N$_{H}^{Gal}$,
inconsistent with the value of $7.8\times10^{20}$ cm$^{-2}$  estimated from HI observations in this direction \citep{kalberla05}.
Allowing the  metallicity in the host absorption to vary, resulted in  
metals abundances  $\lsim$10\%  Solar and   N$_{H}^{Gal}$=(8.5$^{+0.8}_{-1.6}$)$\times10^{20}$ cm$^{-2}$,
fully consistent with the Galactic value.
We therefore fixed N$_{H}^{Gal}$ = $7.8\times10^{20}$ cm$^{-2}$  and obtained the following best fit 
parameters:  power-law photon index $\Gamma$=1.87$\pm$0.01,  
unabsorbed 0.3-10 keV flux  F = (8.7$\pm$0.1)$\times10^{-12}$ erg cm$^{-2}$ s$^{-1}$,  
N$_{H}^{Host}$=(5.1$^{+0.5}_{-0.7}$)$\times10^{22}$ cm$^{-2}$, and   metallicity $<$0.05 Solar
(reduced $\chi^2$=1.13 for 154 d.o.f.).
}

\citet{hur08} compared  different approaches commonly used to search for lines in X-ray spectra. They showed that the highest sensitivity
is obtained with the analysis based on Bayesian posterior predictive probabilities  \citep{pro02} or with the matched filter smoothing
method \citep{rut03}.
In our analysis of the afterglow spectrum of \grb , we implemented   the matched filter smoothing algorithm as follows.

The background-subtracted counts spectrum was convolved with a Gaussian with energy-dependent width $\sigma(E)$.
To properly take into account the instrument energy resolution at the time of our observation,  we derived $\sigma(E)$ by fitting the response matrix produced with the SAS task \textsc{rmfgen}.
The resulting values of  $\sigma(E)$  are well approximated by  the relation $\sigma(E)$ = a + bE + cE$^2$,  with a=0.0422, b=0.0100,  c=--0.000535 and the energy E  in keV.
These values  differ from those used in \cite{rut03} due to the degradation of the CCD performances with time, which resulted in a $\sim20\%$ worse energy resolution.
The Gaussian-convolved spectrum  is shown in Fig.~\ref{fig_spectrum}.

To establish  the threshold values to use in the line search we simulated 10$^5$ spectra with the best fit continuum model described above and convolved them in the same way used for the \grb\ data.
The distributions of the resulting number of counts in each energy channel were then used to derive the percentile curves shown by the dashed lines in Fig.~\ref{fig_spectrum}, which correspond to  single trial detection significances of  2, 3 and 4 $\sigma$.
The convolved data and threshold levels, divided by the model for a better  visualization, are shown in the bottom panel of the figure.

\begin{figure}
\includegraphics[width=85mm]{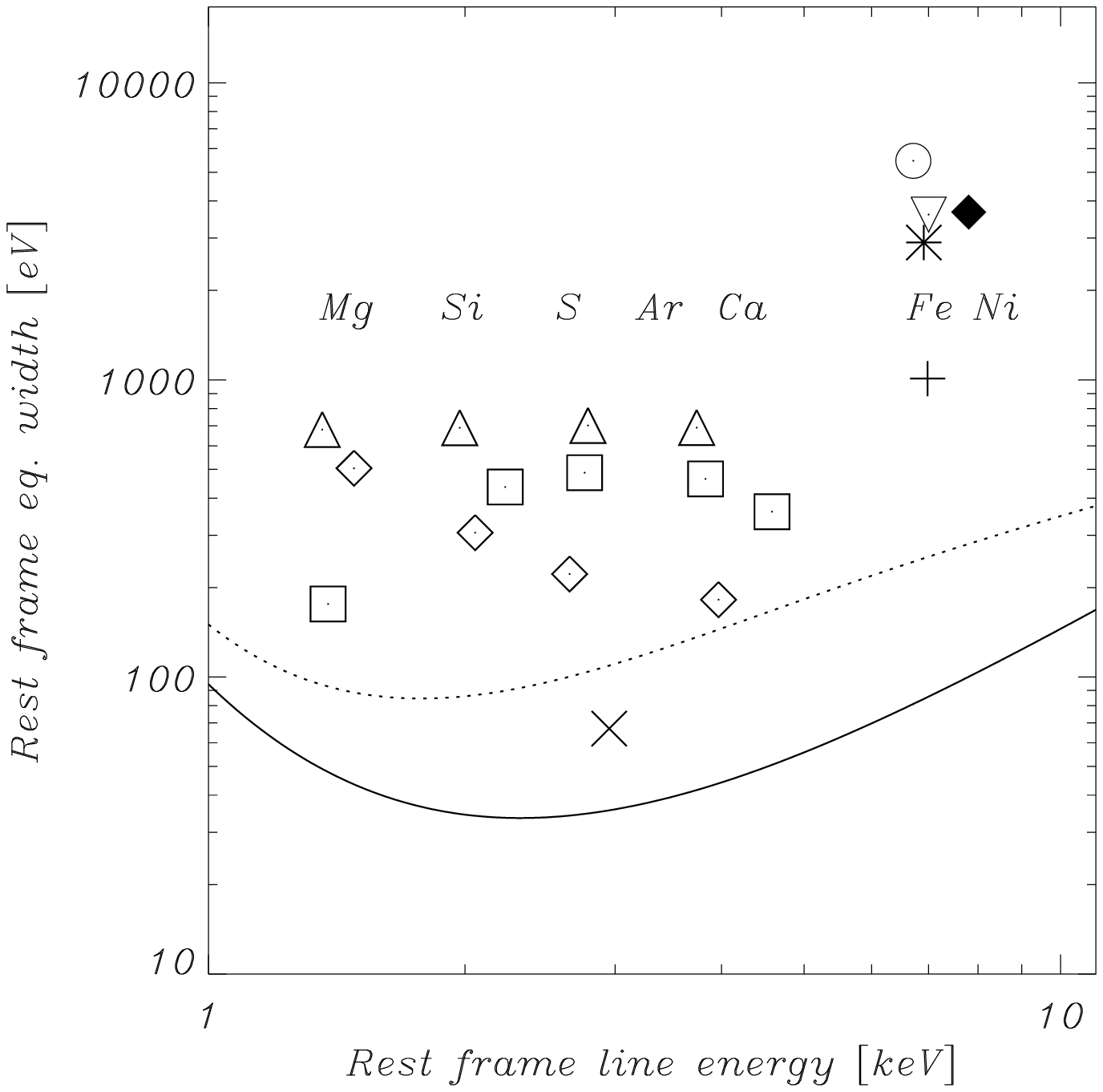}
\caption{\footnotesize {Upper limits on the equivalent width of emission lines in the afterglow of \grb\ in the case of narrow
($\Delta$ E=0 eV, solid curve) or wide ($\Delta$E = 200 eV, dotted curve) lines. 
The curves give the best fit (with a 4th-order logarithmic polynomial) to the values obtained from simulations.
For comparison, also the rest-frame equivalent widths reported for other GRBs are plotted:
$\bigcirc$ GRB 970828,
$\bigtriangledown$  GRB 970508,
$+$ GRB 991216, %
$*$ GRB 000214,%
$\triangle$ GRB 001025,%
squares  GRB 010220,%
filled diamond GRB 011211%
$\times$ GRB 020813%
$\diamondsuit$ GRB 030227%
The energies corresponding to the transition n=2$\rightarrow$1 for some H-like ions are also indicated.
}
 \label{fig_ul}}
\end{figure}

\begin{figure}
\includegraphics[width=85mm]{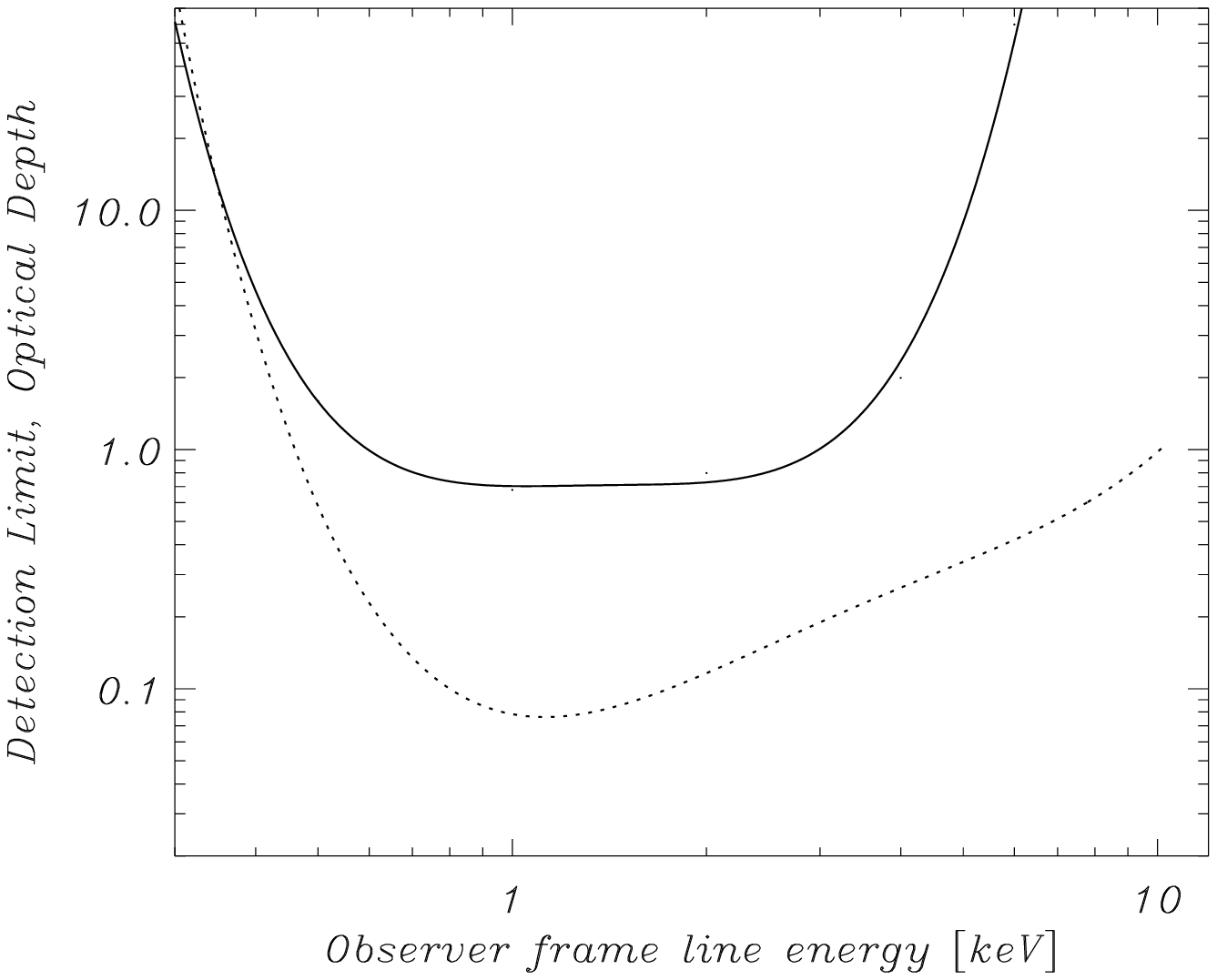}
\caption{\footnotesize {Upper limits on the optical depth  at line center  of absorption lines in the afterglow of \grb\ in the case of narrow
($\Delta$ E=0 eV, solid curve) or wide ($\Delta$E = 200 eV, dotted curve) lines. The curves give the best fit (with a 4th-order logarithmic polynomial) to the values obtained from simulations.}
 \label{fig_ll}}
\end{figure}

The largest deviations in the observed spectrum, occurring at $\sim$6.80 and $\sim$8.23 keV, reach only the 3 $\sigma$ confidence level.
Given  that there is no a-priori reason to expect a line at any of these energies, we have to consider the number of independent trials, which can be taken as the number of resolution elements in the explored energy range, $\sim$ 120, which makes the detection not significant.
 {We repeated the analysis using the model with parameters fixed at the values corresponding to their 2$\sigma$ error regions.
Also in this case no significant features were found.}
We therefore conclude that there is  no evidence for emission or absorption lines in  the EPIC/pn spectrum   of the \grb\ afterglow.

To estimate  the upper limits on emission lines we simulated spectra of \grb\ with Gaussian lines of different widths $\Delta$E, centroid energy,
and intensity  added  to the continuum model.
In this way we determined the line intensities that would exceed our 4$\sigma$ threshold in more than 90\% of the cases.
These intensities are plotted in Fig.~\ref{fig_ul} for the cases of  $\Delta$E=0  (i.e. non-resolved lines)  and $\Delta$E=200 eV  and the best fit model with z= 1.405.
The value of $\Delta$E=200 eV has been chosen as representative for the lines previously reported in the literature \cite[see][]{sak05}.
As done for the emission lines, the limits for the lines in absorption were computed assuming Gaussian profiles,  with widths of 0 eV and 200 eV.
These limits are shown in Fig.~\ref{fig_ll} in terms of optical depth at line center.

 {Two further \xmm\ observations of \grb\ were performed on 2012 July 28-29 and  August 15. 
Although they provide a lower sensitivity compared to the first observation, due to the 
source fainter flux and shorter exposure time, we analysed also these data to 
search for emission lines, possibly appearing at a late time. None was  found, with equivalent width upper limits 
about one order of magnitude higher  than those of Fig.~\ref{fig_ul}.
}

\section{Discussion and conclusions}

Thanks to the large number of counts collected in the afterglow spectrum of \grb\ with the \xmm\ EPIC instrument we have performed a search for emission and absorption lines with unprecedented sensitivity.
The EPIC/pn spectrum we analyzed contained $\sim$ 50000 net counts.
For comparison, the afterglow spectrum with the highest statistics previously available was GRB 040106  \cite[$\sim$23,000 net counts in the EPIC pn plus MOS cameras,][]{sak05}.
A few features were found in its spectrum, of which the brightest was an emission line at $\sim$0.66 keV with equivalent width EW=39 eV.
However these authors estimated that the significance of these lines was  too small to claim a detection.

 {The equivalent widths of some of the lines reported in the literature for other GRBs are plotted in Fig.~\ref{fig_ul} for
comparison with our upper limits.}
We only considered the bursts for which lines were claimed with  statistical significance greater than 2 $\sigma$, and plotted in the figure the measured equivalent width in the rest-frame of the bursts.
For
GRB 970508 \citep{pir99},
GRB 970828 \citep{yoshida01},
GRB 991216 \citep{piro00},
GRB 000214 \citep{antonelli00}, and
GRB 030227 \citep{mereghetti03}
the reported lines were attributed to highly ionized iron, while lines from H- or He-like ions from lighter elements (Mg to Ca) were claimed in
 GRB 001025 \citep{watson02},
GRB 011211 \citep{reeves03},
GRB 020813 \citep{butler03}, and GRB 030227  \citep{watson03}.
A possible line in   GRB 010220 was  attributed to Ni \citep{watson02}.

The equivalent widths reported for these lines are all well above our upper limits, and would hence be largely visible in the afterglow spectrum of \grb.
Our results hence indicate that similar lines are, at least, not common in the spectra of GRBs afterglows.

 {
We also found evidence for a low-metallicity absorbing  medium in the GRB host galaxy. 
This result supports the suggestion that the  high absorption in   X-ray afterglows might  be mainly due to helium in the
HII regions in which the GRBs were  born \citep{wat13}. 
}

\begin{acknowledgements}

This research is based on data of XMM-Newton, an ESA science mission with instruments and contributions directly funded by ESA Member States and NASA. We thank N.Schartel and the staff at the XMM-Newton Science Operation Center for making this target of opportunity observation and A.Tiengo  for advice on the data analysis.
This work was partially suported by the ASI/INAF agreement I/033/10/0.

\end{acknowledgements}


\end{document}